\documentstyle[eqsecnum,floats,prd,aps]{revtex}
\input psfig.tex
\begin{document}
\draft
\newcommand{\be}{\begin{equation}}
\newcommand{\ba}{\begin{eqnarray}}
\newcommand{\ee}{\end{equation}}
\newcommand{\ea}{\end{eqnarray}}

\title{Is The Universe Infinite Or Is It Just Really Big?}
\author{Janna Levin${}$,  
Evan Scannapieco${}$ and 
Joseph Silk${}$}
\address{Center for Particle Astrophysics,
UC Berkeley, Berkeley, CA 94720-7304}
\twocolumn[
\maketitle
\widetext
\begin{abstract}

The global geometry of the universe is in principle as observable an
attribute as local curvature.  Previous studies have
established that if the universe is wrapped into a flat
hypertorus, the simplest compact space, then the fundamental domain
must be at least $0.4$ times the diameter of the observable universe.
Despite a standard lore that the other five compact, orientable flat spaces
are more weakly constrained, we find the same bound holds 
for all.  Our analysis provides the first limits on compact
cosmologies built from the identifications of hexagonal prisms.

\end{abstract}
]
\narrowtext
\begin{picture}(0,0)
\put(410,190){{ CfPA-98-TH-01}}
\end{picture} \vspace*{-0.15 in}

\setcounter{section}{1}

Our universe appears to stretch at least ten billion light years 
across.
As far as the eye can see, there is no visible bound to spacetime.
Still the universe may not be infinite.  It may be more
natural for space to be topologically
compact and multiconnected.
There was once a cultural prejudice that the Earth was flat and unconnected,
so much so that explorers were feared to have fallen off the edge.
The assumption that space must be infinite may represent a similar
bias.
Just as most have realized that our planet is compact, we 
may someday learn that the entire universe is likewise compact
and connected.

Interest in 
compact universes\cite{{lbbs},{sss},{lum},{bl},{css},{bps}}
has spawned several approaches to the search for global topology.
In this {\it letter} we analyze temperature 
fluctuations in the cosmic background radiation (CBR) for all six
compact,
orientable flat topologies.
The simplest compact flat space, the hypertorus, 
was studied in Ref. \cite{sss}.  
Long-wavelength fluctuations could not fit inside a small torus.
The resultant cutoff
in the spectrum of fluctuations was used to bound
the topology scale. 
We show that all six orientable, compact, flat spaces are cutoff at the same 
wavelength
as the hypertorus.
Since the observed quadrupole is in fact low, this alone does not lead to 
a particularly severe bound.  Minimizing the variance of the angular
power spectrum with respect to the COBE data,
we maintain the conservative conclusion that if the universe
is finite and flat, it is bigger than 40\% of the diameter of 
the surface of last scatter. 
The physical universe is then $\ge 2400{(H_o/100{\rm km/s/Mpc})}^{-1}$ 
Mpc across.
Note that
there could still be as many as eight
copies of our cosmos within the observable horizon.


Tiny ripples in the gravitational potential $\Phi $ induce temperature
fluctuations in the CBR via the Sachs-Wolfe effect
	\be
	{\delta T\over T}(\hat n)={1\over 3}\Phi(\Delta \eta \hat n)
	\ee
where $\Delta \eta $ is the conformal time between today and the time of 
decoupling.  The potential can be decomposed into eigenmodes
${\Phi}=\int_{-\infty}^{\infty} d^3\vec k
	\hat \Phi_{\vec k}\exp\left (i\Delta \eta {\vec k}\cdot \hat 
	n\right )$.
The $\hat \Phi_{\vec k}$ are primordially seeded 
Gaussian amplitudes
that obey the reality condition
$\hat \Phi_{\vec k}=\hat \Phi_{-\vec k}^*$.
On a compact manifold, the continuous spectrum of eigenvalues, $\vec k$,
becomes discretized.  
In general we write the temperature fluctuation in any compact, flat 
spacetime as 
	\[
	{\delta T\over T}(\hat n)\propto \sum_{-\infty< k_x,k_y,k_z<\infty}
	\hat \Phi_{jwn}
	\exp\left(i\Delta \eta \vec k \cdot \hat n\right )
	 ,\]
up to a normalization.
As a result of the global topology, 
all of these spaces are anisotropic and all except for the 
hypertorus are inhomogeneous.

\begin{figure}
\centerline{\psfig{file=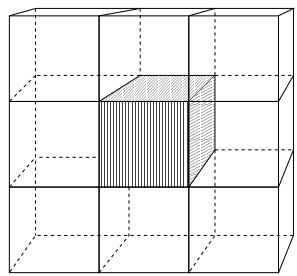,width=1.5in}}
\caption{Tiling flat space with parallelepipeds.}
\label{para}
\end{figure}

\begin{figure}
\centerline{\psfig{file=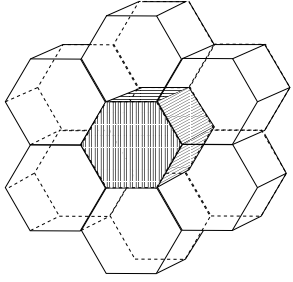,width=1.5in}}
\caption{Tiling flat space with hexagonal prisms.}
\label{hexag}
\end{figure}

Three dimensional Euclidean space can be made topologically compact 
by beginning with either a parallelepiped or a hexagonal prism
as the finite fundamental domain.  Six different orientable, compact
spaces can be constructed by gluing opposite faces of the
fundamental polyhedron.
Another way to represent the identifications is to tile space with 
copies of the fundamental domain as shown in Figs. \ref{para}
and \ref{hexag}.
There are four compact, orientable spaces that can be constructed from 
the parallelepiped tiling and two from the hexagonal tiling.
The hypertorus is the simplest and is built out of a parallelepiped
by identifying $x\rightarrow x+h$, $y\rightarrow y+b$ and $z\rightarrow z+c$.
The identification leads to a restriction of the eigenvalue spectrum,
	\be
	k_x={2\pi \over h} j\quad \quad
	k_y={2\pi \over b} w\quad \quad
	k_z={2\pi \over c} n
	\ \ 
	\ee
with the $j,w,n$ running over all integers.
It is clear that there is a minimum eigenvalue and hence a maximum wavelength
which can fit inside the fundamental domain defined by the parallelepiped
\cite{sss}:
	\[
	k_{\rm min}=2\pi \ {\rm min}\left ( {1\over h},{1\over b},{1\over c}
	\right )\quad 
	{\lambda}_{\rm max} ={\rm max}\left (h,b,c\right)\  .
	\nonumber
	\]
The incompatibility between this truncated angular power spectrum
and a standard flat spectrum means
the width of a square hypertorus had to be $0.8$ the radius of the
surface of last scatter or $0.4$ of the diameter.

Three other spacetimes involve the identification of opposite sides of the
parallelepiped with one or more of the faces twisted before being 
affixed to its counterpart.
It was argued that with these
twists, longer wavelengths could fit in the fundamental domain
since a wave needs to wrap more than once before coming back to 
a fully periodic identification.
However, we find upon closer inspection that these long modes are forbidden
and 
all the compact topologies have the same cutoff
as the hypertorus.

The first twisted parallelepiped we consider 
has opposite faces identified
with one pair rotated through the angle $\pi $.
The periodicity condition requires 
$\Phi(x,y,z)=\Phi(-x,-y,z+c)=\Phi(x,y,z+2c)$.
A translation through $\Delta z=2c$ enforces
	\ba
	{\Phi}(x,y,z)&\equiv
	&\int_{-\infty}^{\infty} 
	\ d^3 \vec k \ \hat \Phi_{\vec k}\ e^{i{\vec k}\cdot 
	\vec x}\nonumber \\
	&=&
	\int_{-\infty}^{\infty} 
	\ d^3 \vec k \ \hat \Phi_{\vec k}\ e^{i{\vec k}\cdot 
	\vec x}e^{ik_z 2c}\nonumber \\
	&=& \Phi(x,y,z+2c)\nonumber \ \ .
	\ea
Matching coefficients, it follows that
$e^{ik_z 2c}=1$.
With the remaining two faces identified without any twists the eigenmodes
are
	\be
	k_x={2\pi \over h}j \quad \quad 	k_y={2\pi \over b} w
	\quad\quad	k_z={\pi\over c}n \ \ .
 \ee
At first glance it appears as though a long mode of wavelength 
$2c$ fits inside the twisted space,
but this eigenmode is disallowed. 
A translation only once through to $\Delta z=c$ requires 
$\Phi(x,y,z)=\Phi(-x,-y,z+c)$ so that	
	\ba
	{\Phi}(x,y,z)&=
	&\sum_{-\infty}^{\infty} 
	\  \hat \Phi_{j w n}\ e^{2\pi i ( (j/h)x+(w/b)y)}e^{i\pi(n/c)z}\nonumber \\
	&=&
	\sum_{-\infty}^{\infty} 
	\ \hat \Phi_{j w n}\ e^{- 2\pi i( (j/h)x+(w/b)y)}
	e^{i \pi {n\over c}(z+c)}
	\nonumber \\
	&=& \Phi(-x,-y,z+c)\nonumber \ \ .
	\ea
Matching coefficients gives
	\be
	\hat \Phi_{jwn}=\hat \Phi_{-j-w n}\ e^{i {\pi}n}
	\label{match}
	\ee
and consequently,
the spectrum is not only
discrete, but there is also a condition 
on the coefficients of the eigenmodes.
As a result,
the cutoff is {\it not} twice as long in the $z-$direction.
If $j=w=0$, then (\ref{match}) requires $n$ to be even and so the lowest 
mode in the $z$-direction is still $k_z=2\pi/c$.  If all scales are set 
equal $j=0,$ and $w=1$, then the $n=1$ mode is allowed but its wavenumber
$k=\sqrt{5}\pi /c$ which is bigger than $k_{\rm min}=2\pi/c$.

\begin{figure}
\centerline{{\psfig{file=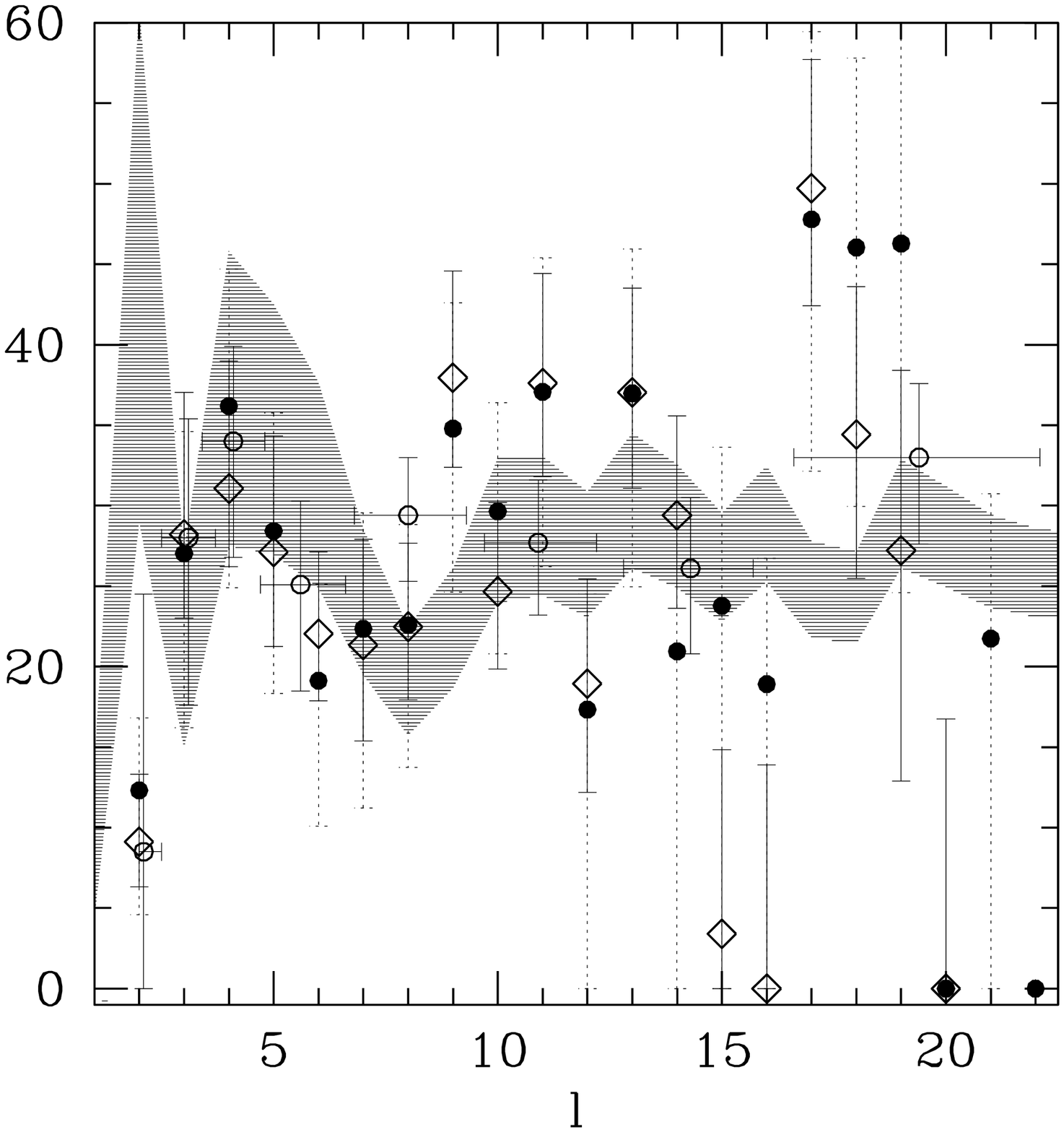,width=1.8in}}
{\psfig{file=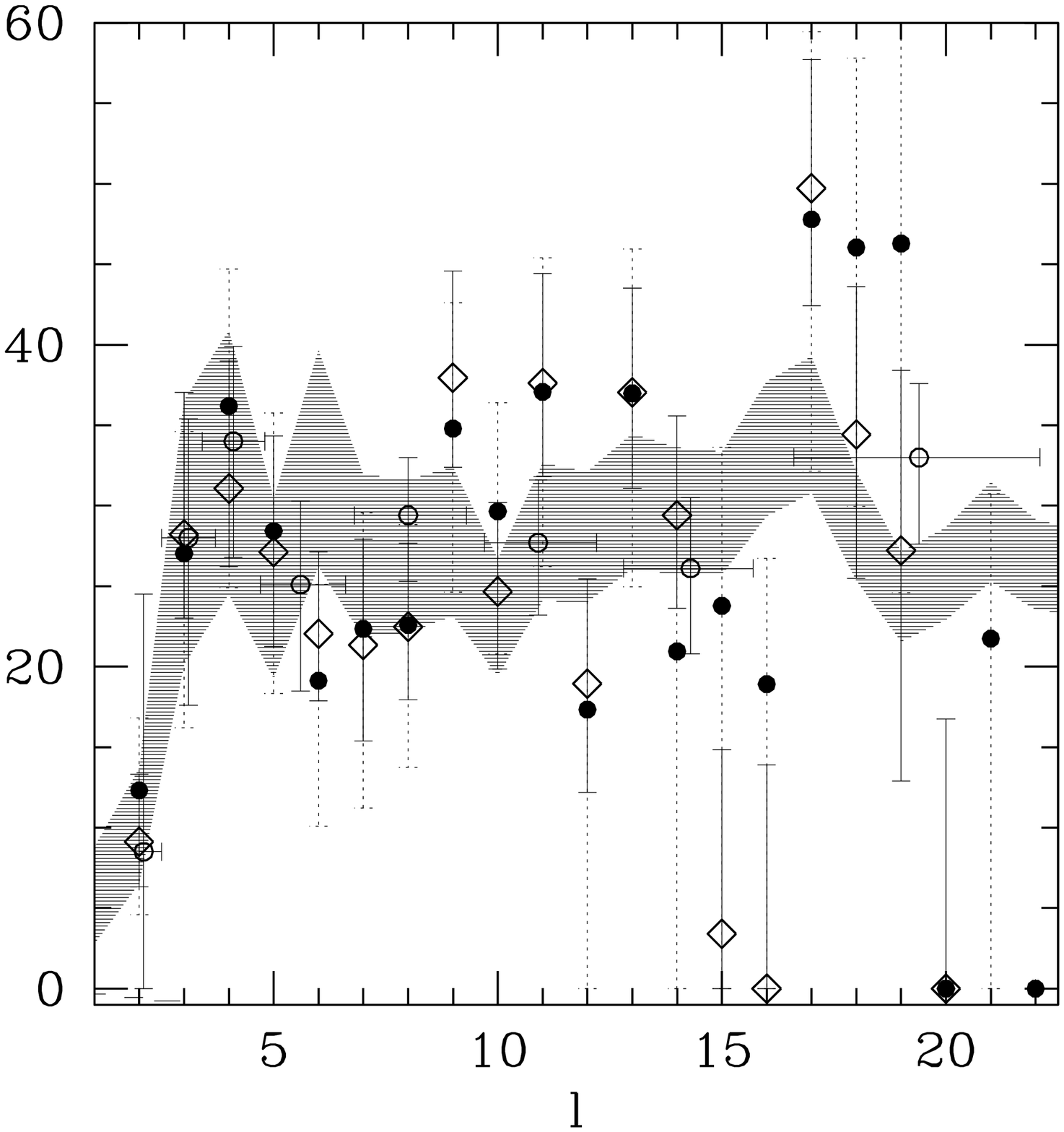,width=1.8in}}}
\caption{
$\delta T/T_\ell$
in $\mu K$ for the hypertorus and 
the $\pi/2$-twisted space.
The shaded band allows for cosmic variance.
The topology scale is equal to
the diameter of the sphere of last scatter with $h=2\Delta \eta$.}
\label{square}
\end{figure}

Another possible compact space identifies opposite faces with 
one
face rotated by $\pi/2$.
After four passes through $z$ one is returned to 
$(x,y,z)=(x,y,z+4c)$
while as before the other faces are identified without twists.
The discrete eigenmodes are
	\be
	k_x={2\pi \over h}j \quad \quad 	k_y={2\pi \over h} w
	\quad \quad k_z={\pi\over 2c}n \ \ . \ee
The width of the fundamental domain along $y$
must equal that along $x$.
The translations along $z$ through $c,2c,$ and $3c$ result
in the following restrictions:
	\ba \nonumber
	\hat \Phi_{jwn}&=&\hat \Phi_{w -j n}e^{in\pi/2}\\
	\nonumber
	&=&\hat \Phi_{-w -jn}e^{in\pi}\\
	\nonumber
	&=&\hat \Phi_{-w j n}e^{i 3n\pi/2}
	\ \ .
	\ea

Again, when $j=w=0$, then $n=4 m$ and the lowest eigenmode has
$k_z=2\pi/c$ for an equal sided parallelepiped.   
Notice also that $\hat \Phi_{01n}=\hat \Phi_{10n}=0 $.
The
$j=w=n=1$ mode is allowed and, setting all
topology scales equal,
$k=3\pi/c$ which is a smaller wavelength.
Statistics 
that average over the sky, such as the angular power spectrum, will
locate a cutoff at roughly the same mode as for the torus.  

The last parallelepiped is a bit more intricate to build.
The identifications are as follows \cite{wolf}:
Translate along $x$ and then 
rotate around $x$ by $\pi$ 
so that $(x,y,z)\rightarrow (x+h,-y,-z)$. 
Next, translate along $y$ and $z$, then rotate around
$y$ by $\pi$
so that $(x,y,z)\rightarrow (-x,y+b,-(z+c))$. 
Finally, translate along $x,y$ and $z$, then rotate around
$z$ by $\pi$
so that $(x,y,z)\rightarrow (-(x+h),-(y+b),z+c)$. 
The discrete spectrum is
	\be
	k_x={\pi \over h }j\quad\quad
	k_y={\pi \over b }w\quad\quad
	k_z={\pi \over c }n\ \ ,
	\ee
with the restricted coefficients
	\ba
	\hat \Phi_{jwn}
	&=&\hat \Phi_{j-w -n}\ e^{i\pi j}\nonumber \\
	&=&\hat \Phi_{-jw -n}\ e^{i\pi(w+n)}\nonumber \\
	&=&\hat \Phi_{-j-w n}\ e^{i\pi(j+w+n)}\ \ .\nonumber	\ea
No permutations of $(0,0,1)$ are allowed although $(0,1,1)$
permutations are accessible, although with fixed phases 
with repect to the fundamental domain.

Spectra for the hypertorus and
the $\pi/2$-twisted parallelepiped, both as big as the observable universe,
are shown in Fig. \ref{square}.
The power spectrum is defined as usual:
$C_{\ell}=\sum_{m}|a_{\ell m}|^2/(2\ell+1)$
with the $a_{\ell m}$ defined by the decomposition of 
the temperature anisotropy into spherical harmonics,
$\delta T/T=\sum_{\ell m}a_{\ell m}Y_{\ell m}$.
For convenience we plot the quantity ${\delta T/T}_\ell
=\left [\ell(\ell +1)C_\ell/2\pi\right ]^{1/2}$ measured in 
$\mu K$ and compare the model values with
the COBE data as analyzed by Gorski (diamonds)
\cite{kris}, by
Tegmark (open circles) \cite{teg}, and by Bond, Jaffe and Knox
(filled circles)
\cite{jaffe}.
While small universes can be ruled out, it is rather fascinating to 
note that these
large cases are marginally consistent with the data.
After all, the observed quadrupole is low.
It is also not
possible to tell if the observed power spectrum is a smooth function or a 
sparcely sampled jagged function.

The hexagonal tiling of flat space
gives two new topologies.
(It is curious to note that flat space can also be tiled with 
fractal hexagons \cite{schro}.)
Three pairs of opposite sides of the hexagon are identified while
in the $z$ direction, the faces are
rotated relative to each other by 
$2\pi/3$.  The potential can be written as
	\ba
	\Phi =&\sum_{n_2 n_3n_z}& \hat \Phi_{n_2 n_3 n_z}
	e^{ik_zn_z}	 
	\times \label{hex}\\
	&\exp &{\left [
	i{2\pi \over h}\left [
	n_2\left ( -x +{1\over \sqrt{3}} y \right )
	+n_3\left ( x +{1\over \sqrt{3}} y \right )\right ] \right ]}
	\nonumber
	\ea
with the eigenmodes
	\be
	k_x={2\pi \over h}j \quad \quad 	k_y={2\pi \over h} w
	\quad \quad 	k_z={2\pi \over 3 c}{n_z}
	\ee
and
	\ba
	\hat \Phi_{n_2, n_3, n_z}
	&=&\hat \Phi_{n_3, -(n_2+n_3), n_z} e^{i2\pi n_z/3}\nonumber \\
	&=&\hat \Phi_{-(n_2+n_3), n_3, n_z} e^{i4\pi n_z/3}
	\ \ .
	\ea

\begin{figure}
\centerline{{\psfig{file=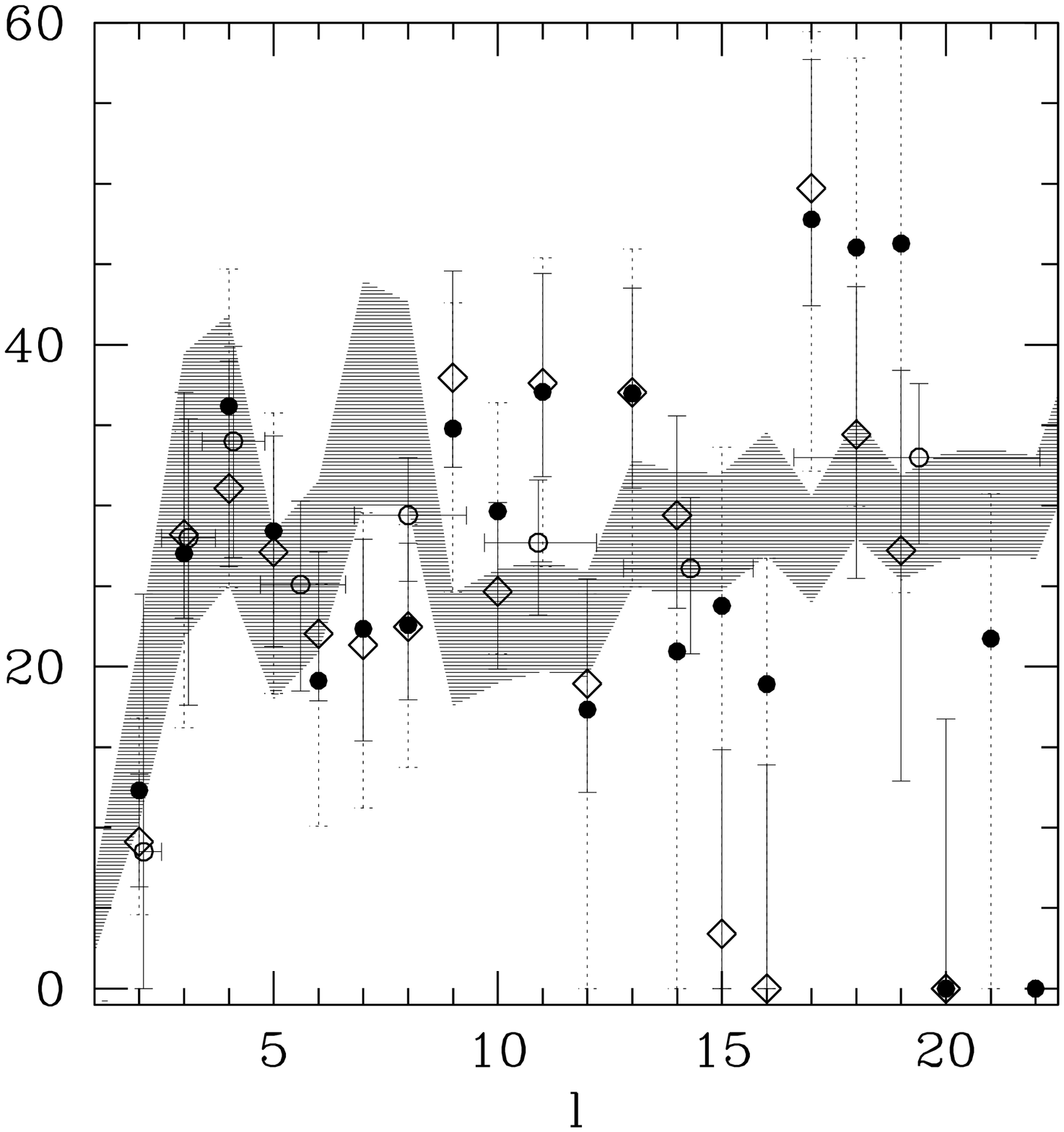,width=1.8in}}
{\psfig{file=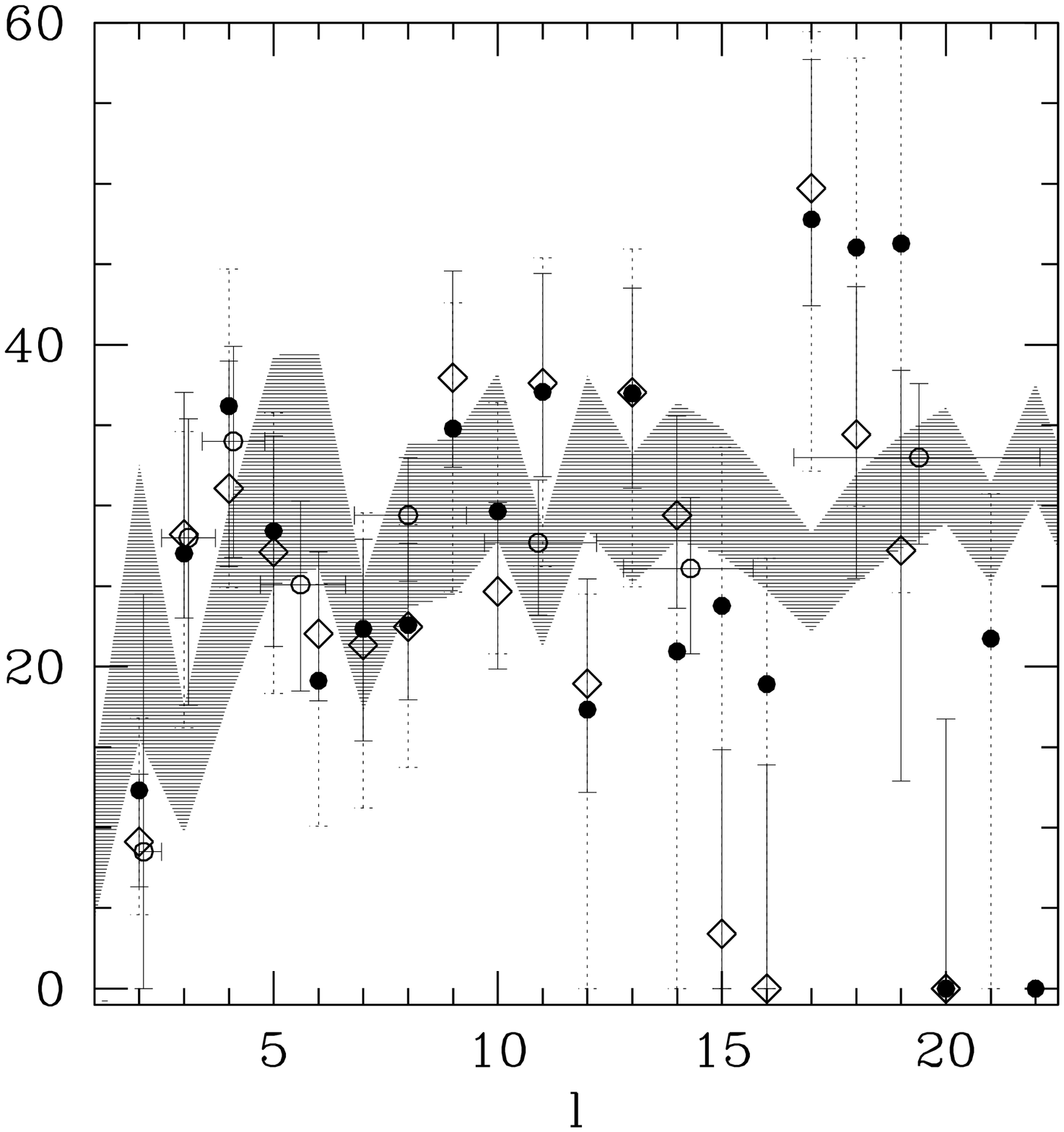,width=1.8in}}}
\caption{${\delta T/T}_\ell$ in $\mu K$
 for the 2 spaces
build from a hexagonal prism
fundamental domain.  The topology scale is comparable
the diameter of the sphere of last scatter with $h=2\Delta \eta$.
The shaded area accounts for cosmic variance.}
\label{big}
\end{figure}

The last possibility 
for 3D Euclidean spaces
identifies the $z$-faces after rotation by $\pi/3$.
The potential 
can still be written as (\ref{hex}) with 
	\be
		k_x={2\pi \over h}j \quad \quad 	k_y={2\pi \over h} w
	\quad \quad k_z={\pi \over 3 c}{n_z}
	\ee
and
	\ba
	\hat \Phi_{n_2, n_3, n_z}
	&=&\hat \Phi_{(n_2+n_3), -(n_2-n_3)/\sqrt{3}, n_z} e^{i\pi n_z/3}
	\nonumber \\
	&=& \hat \Phi_{n_3, -(n_2-n_3), n_z} e^{2i\pi n_z/3}
		\nonumber \\
	&=& \hat \Phi_{-n_2,(n_2-n_3)/\sqrt{3}, n_z} e^{i\pi n_z}
	\nonumber \\
	&=& \hat \Phi_{-(n_2+n_3), n_3, n_z} e^{i4\pi n_z/3}
	\ \ .
	\ea
For both of the hexagons the first $n_2=n_3=0$
mode has $k_{\rm min}=2\pi n_z/c$,
as always.
Spectra for the hexagonal prisms the size of the observable
universe are shown in Fig. \ref{big}.
Again, they are all consistent with the
data.

\begin{figure}
\centerline{{\psfig{file=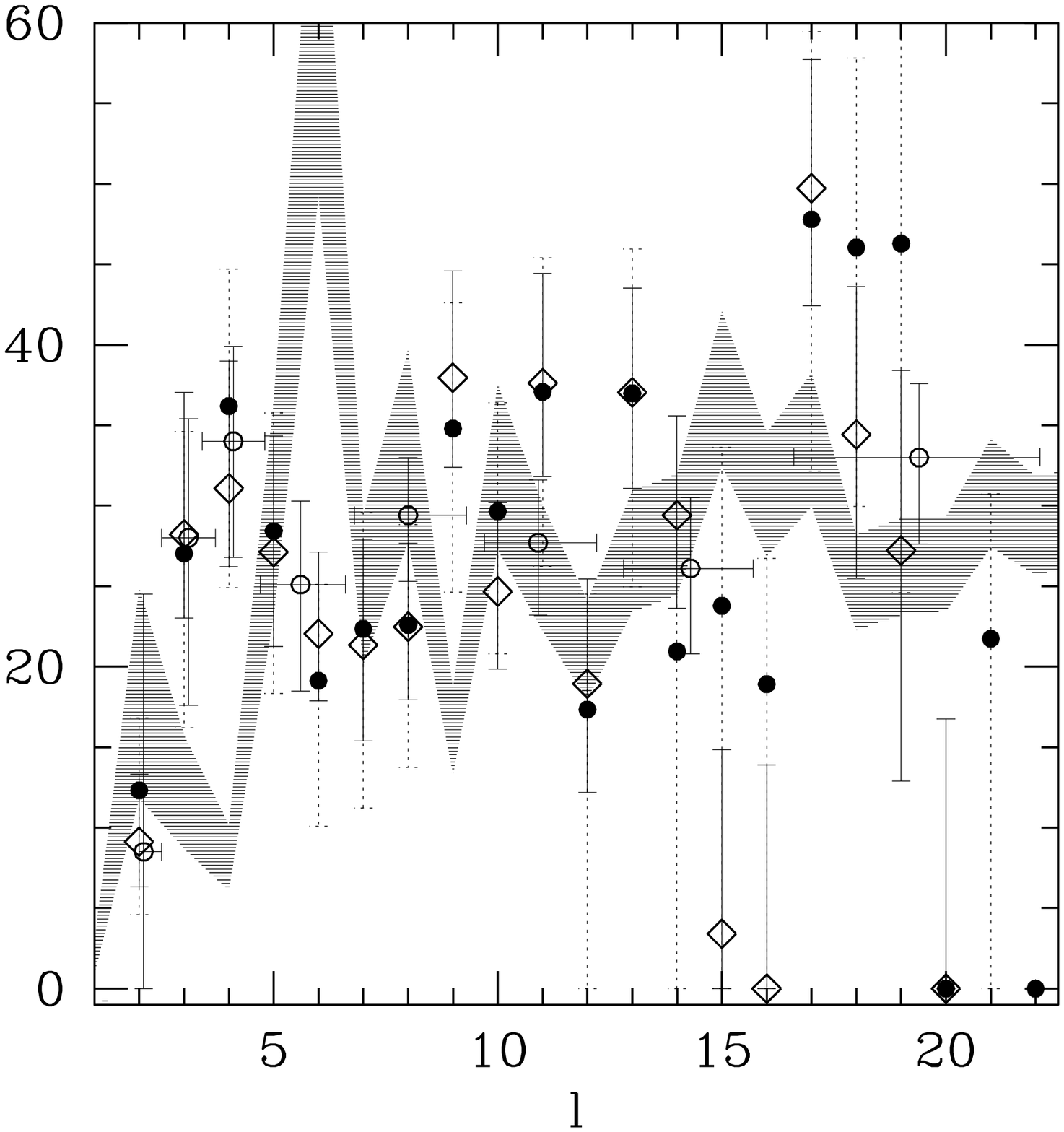,width=1.8in}}
{\psfig{file=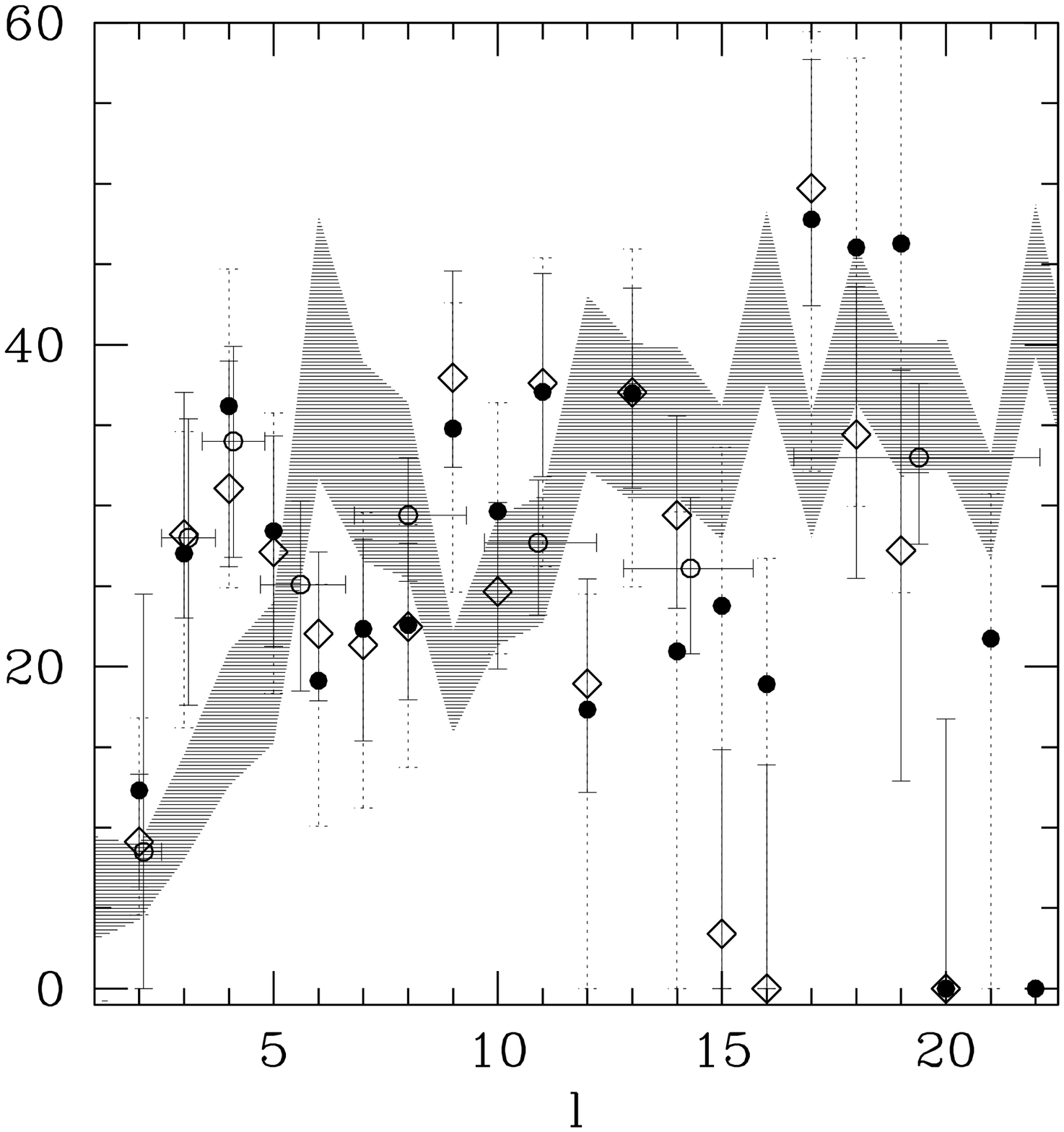,width=1.8in}}}
\caption{
${\delta T/T}_\ell$ in $\mu K$
for two small universes with $h= 0.8 \Delta \eta$.
The leftmost figure is for a regular hypertorus twist and the rightmost
is for the hexagon with a $120^o$ twist.
Cosmic variance is included in the shaded strip.}
\label{small}
\end{figure}

In Fig. \ref{small}, sample small universe spectra are shown.  
As a result of the low observed quadrupole, 
the predicted cutoff alone is not enough to discourage compact models.
We can however ask how likely the spectra shown are relative to 
the spectrum of a flat, infinite cosmology.
Since compact topologies do not give isotropic Gaussian temperature
fluctuations, 
we need to be cautious in making quantitative conclusions based
on any likelihood method.
For this reason, we have maintained the conservative bound
of $h \ge 0.8 \Delta \eta $ which is safely ruled out based on the 
likelihood analysis of
Ref. \cite{jaffe}.
The predicted spectrum is normalized to
COBE by minimizing a variance
in the natural logarithm of
$\delta T/T_\ell$
and the likelihood of the spectrum is assessed given the data.
Compact spaces with a topology scale of $0.8$ 
the radius of the last scattering surface, $0.4$ the diameter,
are at best tens of times less likely than an infinite universe.

\begin{figure}
\centerline{\psfig{file=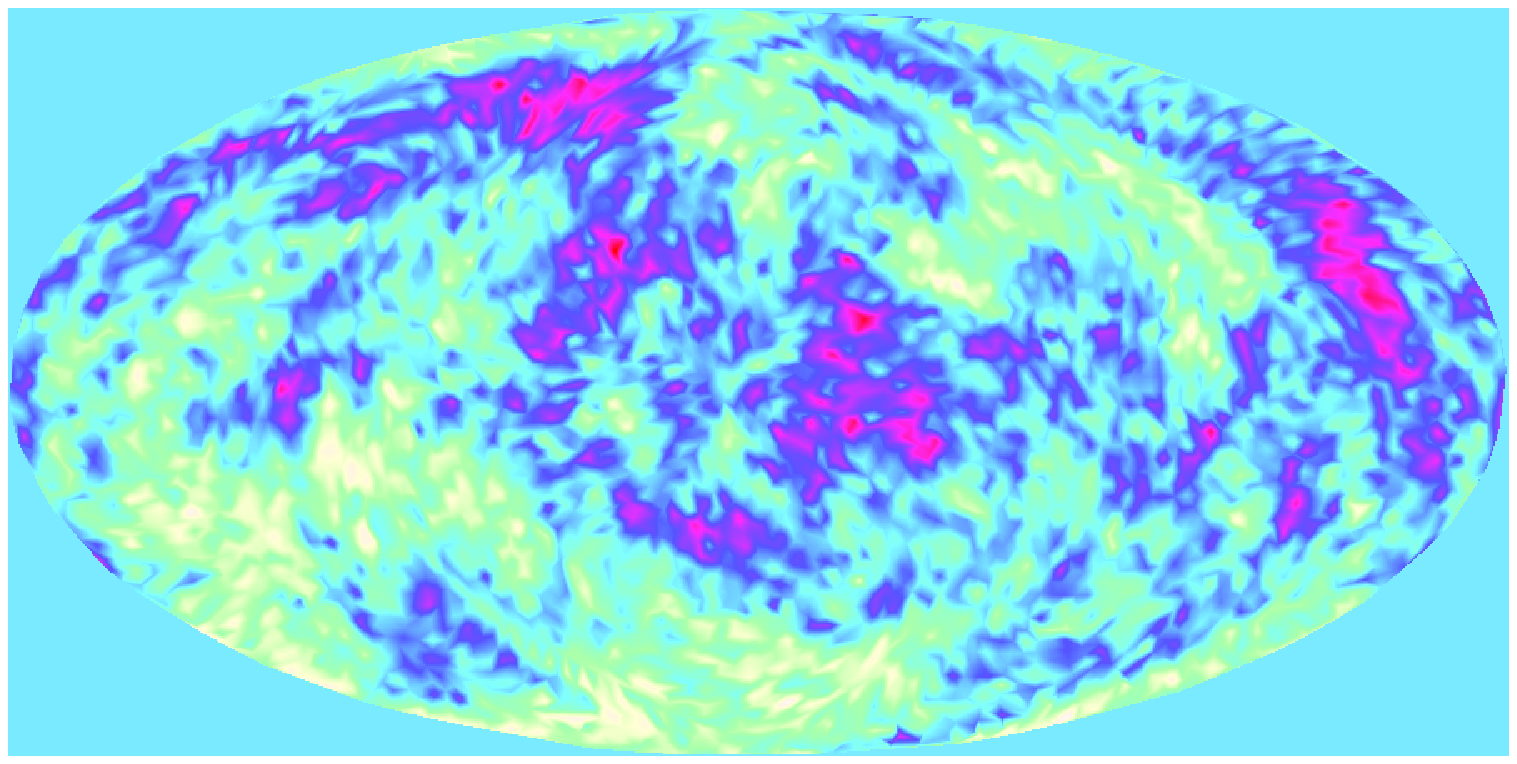,width=3.in}}
\centerline{\psfig{file=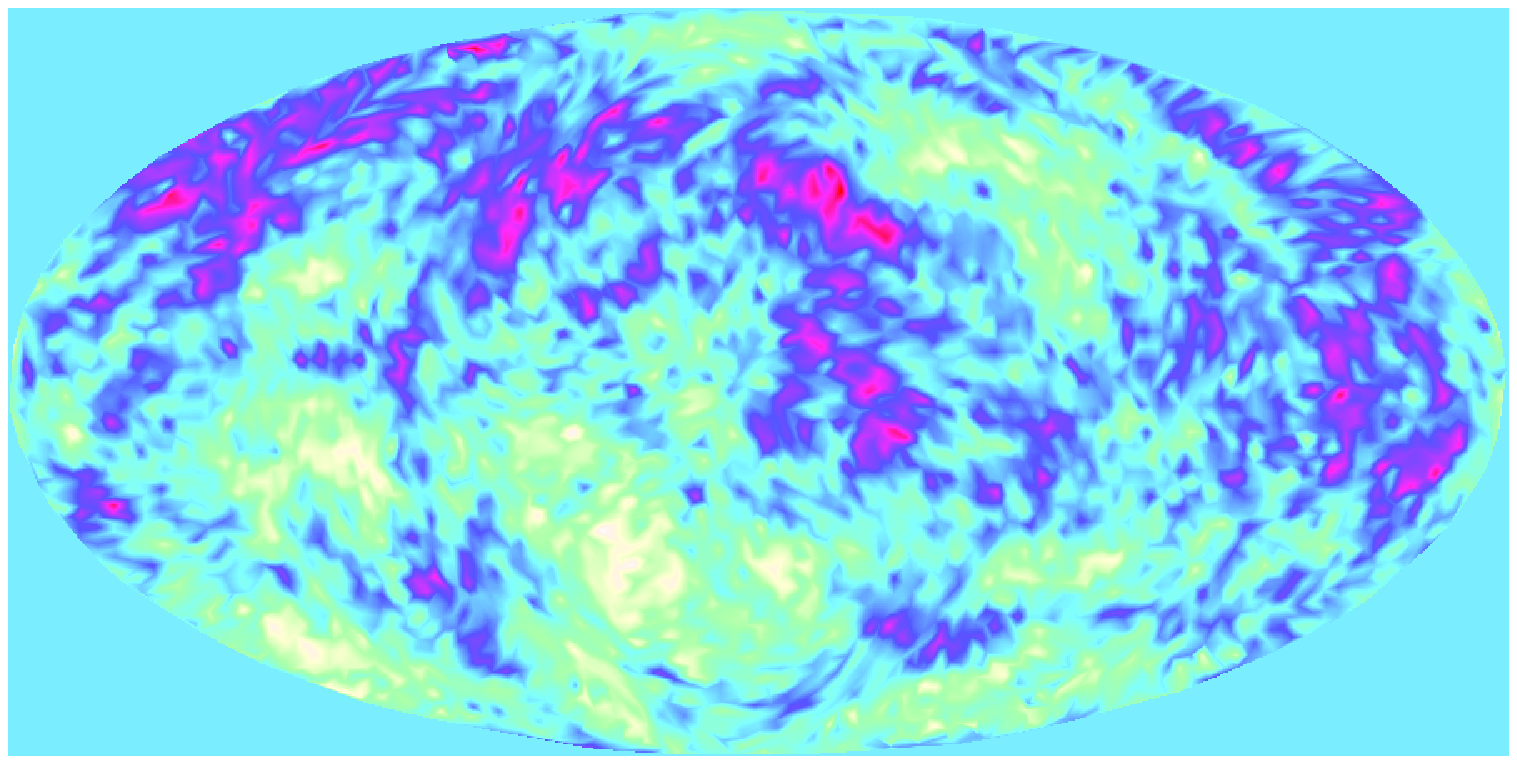,width=3.in}}
\centerline{\psfig{file=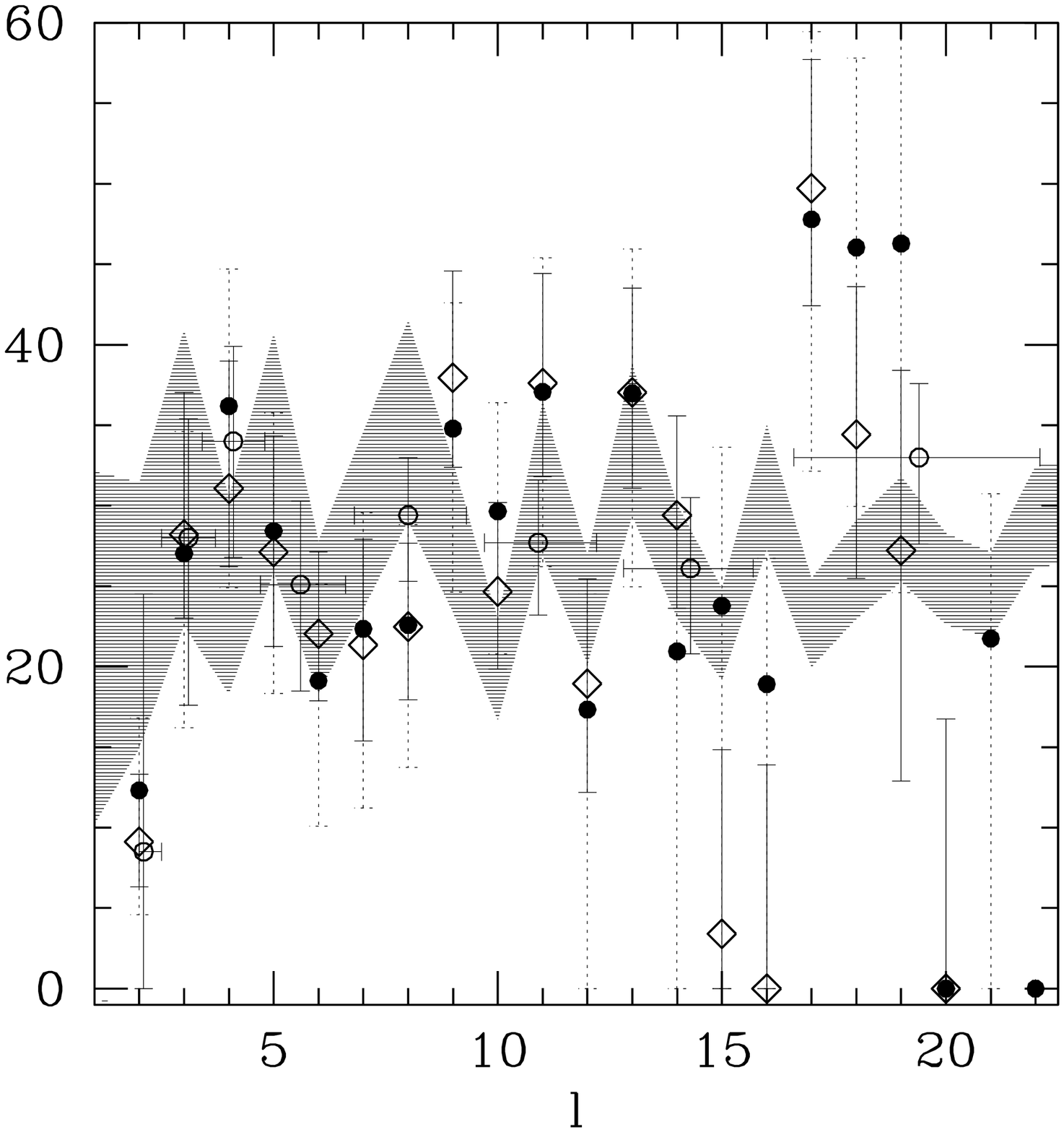,width=1.8in}}
\caption{Top: a simulated map of the microwave sky for a
$\pi$ twisted space with
topology scales 
$h=2,b=1,c=0.5$.  
Middle: same space with the observer offset from 
the origin of the universe.
Bottom: ${\delta T/T}_\ell$ for the lower map.
}
\label{twist}
\end{figure}

While the angular power spectrum
is sufficient to constrain symmetric, flat topology it is in general a poor
discriminant. The average over the sky fails to recognize the strong
inhomogeneity and anisotropy manifest in these cosmologies.
Fig. \ref{twist} 
shows a simulated CBR map of a $\pi$-twisted compact cosmos.  
The topology scales relative to
the  radius of the last scattering surface are 
$h=2,b=1,c=0.5$.  
In the upper panel, the observer
is at the origin.  
In the bottom map, we have offset the observer from the origin.
While the ${\delta T/T}_\ell$ 
can be used to bound equal sided small universes,
the characterization falters for 
this anisotropic example.  The map on the other hand 
shows the markings of the geometry, and a better statistic for extracting
correlations is sorely needed.  
The symmetry studies of Ref. \cite{costa}
that place lower bounds on anisotropic hypertori 
hint at the power of such a pattern-oriented 
approach.
We are currently
developing more powerful and general methods 
to identify patterns hidden in the data
\cite{lssb}.

Our prejudice that the universe is infinite may be no more justified than
the old prejudice that the 
Earth is flat.  
The exploration of the universe for signs of topology is really
just beginning.
While very small flat spaces appear less likely, there are 
an infinite number of compact, negatively curved possibilities
which remain unconstrained.
The search for patterns in sky maps
may be of particular importance
in identifying these \cite{lssb}.
As our vision improves with the aid of future satellite
missions, 
we may soon be able to see, literally see,
the global geometry of our universe.

\begin{center}
\_\_\_\_\_\_\_\_\_\_\_\_\_\_\_\_\_\_\_\_\_\_\_\_\_\_\_
\end{center}

\vskip10truept

We appreciate the valuable input from 
J.R. Bond, N. Cornish, P. Ferreira, K. Gorski, T. Saurodeep,
D. Spergel and G. Starkman.  We are indebted to Andrew Jaffe for 
teaching us 
the likelihood methods of Ref. \cite{jaffe}.
We are also grateful to
Ted Bunn for allowing us to adapt his codes.
E.S. is supported by the NSF.


\begin{references}


\bibitem{lbbs}  J. Levin, J.D. Barrow, E.F. Bunn and J. Silk,
{\em Phys. Rev. Lett.} {\bf 79} (1997) 974 .

\bibitem{sss}  D. Stevens, D. Scott and J. Silk, {\em Phys. Rev. Lett.} {\bf 
71} (1993) 20.

\bibitem{lum} M. Lachieze-Rey and J.P. Luminet, {\em Phys. Rep.} {\bf 254}
(1995) 135;
R. Lehoucq, M. Lachieze-Rey and J.P. Luminet,
``Cosmic Crystallography'', gr-qc/9604050.

\bibitem{bl}  J.D. Barrow and J. Levin,{\em Phys. Lett.}
A {\bf 233} (1997) 169.


\bibitem{css} N.J. Cornish, D. Spergel and G. Starkman, 
astro-ph/9602039 (1996); {\it ibid.}
astro-ph/9708225
(1997).

\bibitem{bps}  J. R. Bond, D. Pogosyan and T. Souradeep, preprint
astro-ph/9702212 (1997); {\it ibid.} in preparation.


\bibitem{wolf} J.A. Wolf, ``Spaces of Constant Curvature''
(Publish or Perish, Inc., Wilmington, Delaware, 1967).

\bibitem{kris} K. Gorski, 
Proc. Moriond XVI, 
ed. F.R. Bouchet et. al. (Gif-Sur-Yvette:  Editions Fronti\`ers) (1997).

\bibitem{teg} M. Tegmark {\em Phys. Rev.}
{\bf D 55} (1997) 5895.




\bibitem{jaffe} J.R. Bond, A. Jaffe and L. Knox,
preprint CfPA-97-TH-11; {\it ibid.} in preparation;
J.R. Bond and A. Jaffe,
Proc. Moriond XVI, 
ed. F.R. Bouchet et. al. (Gif-Sur-Yvette:  Editions Fronti\`ers) (1997).


\bibitem{schro} Schr\"oder, {\it Chaos, Fractals and Power Laws}
(W.H.Freeman and Company, 1991).

\bibitem{costa} 
A. de Oliveira-Costa and G. F. Smoot,
Ap.J. {\bf 448} (1995) 447;
A. de Oliveira-Costa, G. F. Smoot, and
A. A. Starobinsky, Ap. J. {\bf 468} (1996) 457.

\bibitem{lssb}  J. Levin, E. Scannapieco, J. Silk and J.D. Barrow,
``How the Universe Got Its Spots'', in preparation.


\end{references}
\end{document}